# ROLE OF CENTRAL BANK IN ADVANCING SUSTAINABLE FINANCE

**Authors**

**A T M Omor Faruq**, Graduate Student, Western Illinois University, Formal Deputy Director, Bangladesh Bank, Central Bank of Bangladesh. Email: ao-faruq@wiu.edu
Address: 617 Reeveston Dr, Macomb, IL 61455, USA

**Md Toufiqul Huq**, Joint Director, Bangladesh Bank, Central Bank of Bangladesh, Upcoming Graduate Student, North Carolina State University, USA. Email: sowrov07@gmail.com
Address: 617 Reeveston Dr, Macomb, IL 61455, USA

## ABSTRACT

This paper examines the pivotal role central banks play in advancing sustainable finance, a crucial component in addressing global environmental and social challenges. As supervisors of financial stability and economic growth, central banks have dominance over the financial system to influence how a country moves towards sustainable economy. The chapter explores how central banks integrate sustainability into their monetary policies, regulatory frameworks, and financial market operations. It highlights the ways in which central banks can promote green finance through sustainable investment principles, climate risk assessments, and green bond markets. Additionally, the chapter examines the collaborative efforts between central banks, governments, and international institutions to align financial systems with sustainability goals. By investigating case studies and best practices, the chapter provides a comprehensive understanding of the strategies central banks employ to foster a resilient and sustainable financial landscape. The findings underscore the imperative for central banks to balance traditional mandates with the emerging necessity to support sustainable development, ultimately contributing to the broader agenda of achieving global sustainability targets.

# Contents

# 1. Introduction

## 1.1 Context and Importance

The world now faces unprecedent growing environmental and societal challenges, suggesting a need for financial markets to move towards sustainability. Sustainable finance seeks to act in support of economic development without compromising the protection of our environment or social welfare and thereby, aiming at securing sustainable goals for everyone (UN SDGs). This shift is motivated by concerns that, in general, financial systems are already insufficient to deal with the urgent long-term challenges of climate change, biodiversity loss and social justice when regular economic interests prioritize short term benefit over systemic stability.

Given that central banks are the most important players in terms of how we create and regulate money, they have a unique role to play here. Such was the role of a central bank, which primarily focused on financial and macroeconomic stability: but increasingly these guardians are adding ethical oversight in their functionaries to promote sustainability that goes together with finance. As regulators, monetary authorities and financial market players themselves they have important capabilities to help advance sustainable finance.

## 1.2 Objectives and Scope of the Chapter

This Chapter explores the role of central banks in greening the financial system. It seeks to establish a thorough picture of how central banks establish or try to incorporate sustainability into their monetary policy and regulatory frameworks, as well as in the sub-sectors that they oversee. The chapter will examine specific tools and strategies such as sustainable investment principles, climate risk assessments, and the promotion of green bond markets, green financial products etc. Additionally, it will highlight collaborative efforts between central banks, governments, and international institutions to incorporate sustainability into the financial system. By discussing case studies and best practices of different central banks, the chapter focuses on the importance of building a sustainable financial system in addition to traditional central bank mandates.

# 2. Theoretical Framework

## 2.1 Central Bank and its mandates

A central bank is a chief regulator of the financial system with the role of supervising the monetary system and controlling a country's currency, money supply, and interest rates. It acts as a regulator and supervisor of financial services to the government and commercial banks. In any country, a central bank ensures a stable and efficient financial system for economic growth. By setting interest rates thresholds, controlling money flow in market, and conducting open market operations, central banks implement monetary policy to control economic activity, manage inflation, and stabilize the currency. It supervises schedules for banks and other financial institutions as well as extends financial support to the government as a lender. Through this activity, it maintains financial stability of a country. Central banks also issue and circulate the country; s currency, ensuring an adequate supply for economic transactions. Additionally, they manage the country's foreign exchange reserves and influence exchange rate policies to maintain a stable currency value central banks frequently serve as the government's financial representative, overseeing government accounts, issuing public debt, and offering financial guidance. Central banks guarantee secure and reliable financial transactions by overseeing payment systems. The Federal Reserve System, the European Central Bank, the Bank of England, the Bank of Japan, and Bangladesh Bank are few examples of central banks. The independence of the central bank is crucial to maintain long term financial stability and public trust in the financial system.

## 2.2 Understanding Sustainable Finance

Sustainable finance states financial activities or systems that consider environmental, social, and governance (ESG) principles. It aims to promote long term economic growth while addressing climate change, social inequalities, and other sustainability challenges. It supports mobilizing capital flows towards sustainable activities and projects [1]. Over the last few decades, the concept of sustainable finance has developed with growing concerns of the negative impacts resulting from climate change and the potential benefits of sustainable investments. Transparency, accountability, and ESG based financial decision making are few key principles of sustainable finance

This relatively new financial concept can take various forms or instruments for instances green or climate bonds, social impact bonds, sustainable loans, and ESG investment etc. These financial instruments are designed to mobilize funds for projects that have positive environmental and social outcomes. For instance, green bonds are specifically issued to finance projects that reduce carbon emissions or enhance energy efficiency. Social impact bonds, on the other hand, aim to address social issues such as education, healthcare, and housing.

The terms Sustainable Finance, Green Banking, Climate Finance etc. are frequently used interchangeably to indicate the same sort of concepts. However, they perceived to have some key differences. Here we explain some key terms from broader aspects.

### 2.2.1 Sustainable Finance

Sustainable finance implies the procedure of giving environmental, social, and governance (ESG) issues due weight during making investment decisions. In other words, Sustainable finance refers to making investment choices that consider outcomes of an economic action or project through the lens of ESG [2]. The goal of sustainable finance is to channel financing to responsible companies and support the stable financial system in emerging countries by incorporating environmental, social, and governance requirements, along with risk management, into financial institution lending process.

### 2.2.3 Green Banking

The term "green banking" explains the banking industry's initiatives to reduce their carbon footprint and encourage eco-friendly banking practices. The US Environment Protection agency terms Green Banks as institutions that utilize public savings to draw private investments for clean energy projects [3]. This idea includes a variety of actions and regulations meant to promote sustainable development, like investing in eco-friendly projects, putting energy-saving technology into use, and enticing clients to lead eco-friendly lives.

### 2.2.4 Environmental, Social and Governance (ESG)

ESG is a fundamental consideration of Sustainable Finance. Li et al. [4] stated ESG as a philosophy that focuses on long term growth by taking consideration of economic, environmental, social and governance benefits. Environmental factors can include climate change related adaptation and mitigation along with other environmental issues like pollution reduction, biodiversity protection, and the circular economy. Inequality, inclusivity, labor affairs, community & human capital investments, as well as human rights concerns, are all examples of social considerations. Governance factors include board structure and diversity, corporate governance policies, transparency and disclosure, stakeholders' rights etc. The governance of public and commercial institutions plays a major role in determining how social and environmental elements are included in the decision-making process.

### 2.2.5 Climate Finance

Finance that tackles issues linked to loss and damage caused by climate change, as well as that which attempts to lessen the causes of climate disruption and promote adaptation. According to the UNFCCC [5], "Climate Finance refers to national or transnational financing—drawn from public, private and alternative sources of financing—that seeks to support mitigation and adaptation actions that will address climate change".

### 2.2.6 Environmental Finance

Environmental finance is a field of sustainable finance that focuses on investing in, funding, and managing financial resources in a way that supports environmental sustainability. More preciously it exploits various financial tools or instruments to project biodiversity [6]. It encompasses a variety of financial instruments, practices, and strategies aimed at promoting environmentally friendly projects and companies. This includes investments in renewable energy, energy efficiency, water management, sustainable & smart agriculture, and other projects that minimize environmental impact and promote the conservation of natural resources.

### 2.3 Changing role of Central Bank in financial system

As we discussed, Central banks are key players for maintaining financial stability, controlling inflation, and fostering economic growth. They achieve these goals through various tools, including setting interest rates, controlling financial institutions, and injecting liquidity into the banking system. As supervisors of financial stability, central banks have significant control on financial markets and institutions. This influence positions them to promote financial sustainability by adding ESG considerations into their operations and policies [7].

Central banks' traditional mandates often include price stability and full employment. However, there is growing recognition that financial stability cannot be achieved without considering environmental and social risks. For instance, climate change presents systemic threats to the financial system through physical risks, such as extreme weather events, and transition risks, like policy changes designed to cut carbon emissions [8]. By incorporating these risks into their frameworks, central banks can strengthen the financial system.

The traditional understanding of central banking has been changing rapidly to match the growing diversified needs of the modern financial world. The pandemic, climate fall out and other non-traditional threats and risks to financial stability demand attention from centrals banks. Lars Peter Hansen [9] wrote central banks, by using monetary policy, can promote effective methodologies for measuring the enduring effects of climate change exposure. We see there are growing trends of cooperation on sustainability and climate related issues for financial system stability and economic growth among central banks from the developed and developing world as well as other financial sector players. This is tantamount to acknowledging both the threat and potential of climate change risks and sustainable financing and investment. These cooperation among the central banks advocate and encourage integrating climate change and sustainability issues into the so-called traditional central bank policies such as monetary policies, financial stability policies, supervision frameworks, etc.

### 2.4 Nexus between Central Banking and Sustainability

The nexus between central banking and sustainability has become an increasingly prominent area of focus, as central banks recognize their potential role in promoting sustainable economic development. This connection encompasses several dimensions; here are some key aspects of this nexus:

#### 2.4.1 Climate Change and Financial Stability

**Physical Risks:** Central banks are concerned about the financial risks presented by climate related events, like extreme weather, which can impact asset values and financial stability. These are examples of physical risks.

**Transition Risks:** Transition risks refer to the uncertainty associated with shifting to a low-carbon economy, which can affect the value of assets and sectors, potentially causing financial instability if not properly managed.

#### 2.4.2 Green Finance and Investment

Central banks can develop green finance markets by creating favorable regulatory environments, promoting financial instruments such as green bonds, green loan etc. They can also incorporate climate related risks into their stability assessments. They can also invest in sustainable assets and influence financial bodies to adopt ESG factors into their investment strategies.

#### 2.4.3 Monetary Policy and Climate Change

**Monetary Policy Instruments:** To promote and support sustainability central banks can adapt their monetary policy tools. They might incorporate climate risks into their collateral frameworks or consider the environmental impact of the assets they purchase. For example, the Governing Council of the European Central Bank (ECB) has committed to further integrating climate change factors into the Euro system's monetary policy framework. [10].

**Quantitative Easing:** In an effort to encourage sustainable investments, some financial system regulators have begun to merge green bonds into their asset acquisition program.

### 2.4.4 Research and Advocacy

Central banks are conducting research on the economic impacts of climate instability and the effectiveness of various policy measures to mitigate these impacts. They can use their influential positions to advocate for policies that encourage sustainable development goals (SDGs) and to raise awareness of the economic implications of climate disruption.

### 2.4.5 Regulation and Supervision

To make sure that financial institutions are resilient to these risks, monetary regulators can incorporate climate-related risks into their regulation and supervisory frameworks. This entails conducting stress tests on financial institutions to identify climate related risk and verifying that they have implemented sufficient risk management procedures. For instance, In addition to continuously incorporating climate concerns into its regulatory operations, the Bank of Ireland regularly monitors regulated financial market participants' obligations with relation to climate risk and sustainable finance. [11].

### 2.4.6 International Cooperation

Collaboration with other central banks, international organizations, and stakeholders is crucial for addressing global sustainability challenges. The Network for Greening the Financial System (NGFS) is such an organization which assists the sharing of knowledge and best practices for financial regulators throughout the world.

### 2.4.7 Supporting Sustainable Economic Growth

Central banks can support policies that promote sustainable economic growth, which includes encouraging investments in renewable energy sources, energy efficiency, and other sustainable industries. They can also support financial inclusion and access to finance for underserved communities, contributing to broader social sustainability.

## 3. How Central Bank Advancing Sustainable Finance

Central banks are taking a multifaceted approach to advancing sustainable finance. They are integrating sustainability into their core policies, ensuring that climate-related risks are considered in monetary policy and supervisory activities. Promoting a sustainable financial system, central banks support the enhancement of financial instruments that fund environmentally friendly projects. Through collaborative efforts with other financial institutions and stakeholders, they enhance the overall financial system's resilience to climate change impacts. Additionally, central banks are adopting in-house sustainability practices, leading by example in their operations and setting a standard for environmental responsibility within the financial sector.

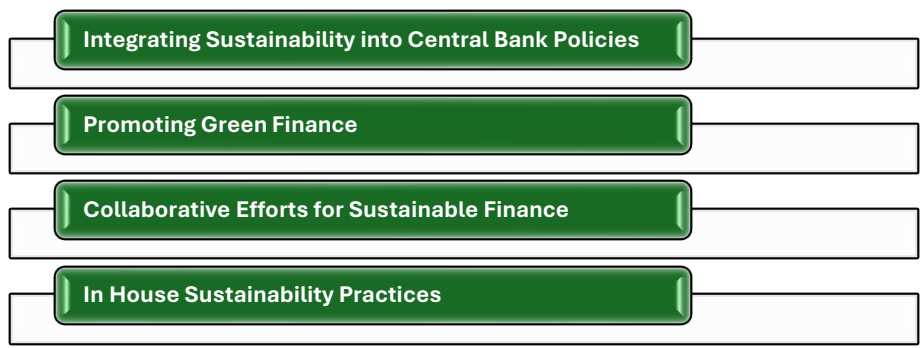

Figure 1: Central Bank Approaches for Integrating Sustainability

### 3.1 Integrating Sustainability into Central Bank Policies

#### 3.1.1 Monetary Policy and Sustainability

Monetary policy traditionally focuses on inflation control and economic progression. However, central banks can also use monetary policy tools to promote sustainability. For example, by considering climate related risks in their economic models and forecasts, central banks can better understand and address the impacts of environmental change on the economy. Additionally, central banks can adopt green quantitative easing (QE) programs, where they purchase green bonds or other sustainable assets to support the development of green finance markets.

The European Central Bank (ECB) is a leading example of adding sustainability into monetary tools. The ECB has included climate related risks in its economic models and has begun purchasing green bonds [12] as part of its asset purchasing programs. This approach not only supports the green bond market but also signals the ECB's commitment to addressing climate change.

#### 3.1.2 Regulatory Frameworks and Supervision

Central banks, often in their capacity as financial regulators, play a key role in ensuring the stability and integrity of the financial system. By integrating sustainability into regulatory frameworks, central banks can promote the adoption of sustainable practices by financial institutions. This can include requiring banks to assess and disclose climate-related risks, encouraging the development of green financial products, and setting standards for sustainable lending practices.

In the UK, the Bank of England has made significant efforts to incorporate climate-related risks into its regulatory system. The BoE requires banks and insurers to assess and report their vulnerability to climate-related risks [13]. It also performs stress testing on climate to assess the strength of the financial system to different climate scenarios [14].

#### 3.1.3 Financial Market Operations

Central banks influence financial markets through various operations, such as open market policy, lending facilities, and reserve requirements. By incorporating sustainability criteria into these operations, central banks can promote green finance and support the transition to a sustainable economy. For instance, central banks can prioritize the purchase of green bonds or provide preferential lending rates to banks that finance sustainable projects.

The People's Bank of China (PBOC) is a pioneer in using financial market operations to promote green finance. The PBOC has implemented green credit guidelines [15] that encourage banks to provide loans for environmentally friendly projects . Additionally, the PBOC has developed a green financial system that includes green bonds, green loans, and green funds, contributing to the rapid growth of China's green finance market.

### 3.2 Promoting Green Finance

#### 3.2.1 Sustainable Investment Principles

Sustainable investment involves incorporating ESG factors into investment decisions to achieve long term financial returns and positive social and environmental impacts. Central banks can promote sustainable investment principles by setting an example with their own investment portfolios and encouraging other financial institutions to do the same. This can involve investing in green bonds, renewable energy projects, and other sustainable assets.

Central banks can also provide guidelines and frameworks for sustainable investment. For example, the Central Bank of Ireland has developed a Sustainable Investment Charter that incorporates how the bank follow sustainable investment principles in its investment practices [16]. By leading by example, central banks can influence other investors to adopt sustainable investment practices.

#### 3.2.2 Climate Risk Assessments

Climate change poses significant risks to the stability of the financial system. Central banks can play a crucial role in assessing these risks and promoting climate resilience within the financial sector. This involves conducting climate stress tests, which assess the impact of different climate scenarios on financial institutions and the economy. Central banks can also develop guidelines and frameworks for financial institutions to assess and disclose their climate-related risks.

The Bank of England (BoE) has been at the forefront of climate risk assessments. It has conducted climate stress tests or Climate Biennial Exploratory Scenario (CBES) to evaluate the impact of climate change and explore the financial risk posed by climate change [17]. The results of these stress tests provide valuable insights into the vulnerabilities of the financial system and help inform policy decisions.

#### 3.2.3 Development of Green Bond Markets

Green bonds are debt instruments specifically designed to fund projects with environmental benefits. They are a key tool in mobilizing capital for sustainable development. Central banks can support the development of green bond markets by purchasing green bonds, setting standards for green bond issuance, and encouraging financial institutions to invest in green bonds. Successful examples of central banks promoting green bond markets include the European Central Bank and the People's Bank of China.

The European Central Bank (ECB) has played a significant role in promoting the green bond market in Europe. By purchasing green bonds as part of its asset purchase programs, the ECB has provided liquidity to the green bond market and encouraged the issuance of more green bonds. This has contributed to the growth of the green bond market in Europe and has supported the financing of environmentally friendly projects.

### 3.3 Collaborative Efforts for Sustainable Finance

### 3.3.1 Collaboration with Governments

Central banks can enhance their efforts to promote sustainable finance through collaboration with governments. This can involve joint policy initiatives, coordinated regulatory frameworks, and shared goals for sustainable development. For instance, central banks and governments can work together to develop national green finance strategies, promote green public procurement, and support sustainable infrastructure projects.

In Germany, the Deutsche Bundesbank collaborates closely with the government to promote sustainable finance. This includes working on national strategies for green finance and supporting government initiatives to enhance sustainability in the financial sector [18]. Such collaboration ensures that policies are coherent and mutually reinforcing.

### 3.3.2 International Institutions and Frameworks

Global challenges require coordinated global responses. Central banks can engage with international institutions such as the International Monetary Fund (IMF), the World Bank, and the Network for Greening the Financial System (NGFS) to promote sustainable finance. These collaborations can involve sharing best practices, developing global standards, and supporting international sustainability initiatives.

The Network for Greening the Financial System (NGFS) is a notable example of international collaboration. It comprises central banks and supervisors from around the world who work together to enhance the role of the financial system in managing climate and environmental risks [19]. By participating in the NGFS, central banks can share knowledge and experiences, which helps in developing effective strategies for promoting sustainable finance.

### 3.4 In house ESG Practices in Central Bank

Central banks are increasingly integrating ESG (Environmental, Social, and Governance) practices into their internal operations, demonstrating commitment to sustainability and responsible governance:

### 3. 4.1 Environmental Practices

**Energy Efficiency:** Central banks upgrade lighting, HVAC systems, and building designs to reduce energy consumption and carbon footprints. Example: The European Central Bank announced to reduce its electricity consumption in main building by 3% by 2023 with 2018 consumption taken as baseline [20]

**Renewable Energy:** Adopting renewable energy sources to power their facilities, reducing reliance on fossil fuels. For example, Bangladesh Bank, Central Bank of Bangladesh, put solar power systems on its rooftop in 2011 to be more energy efficient. Since then, other institutions have also incorporated this technology into their operations [21]

**Green Building Standards:** Designing and maintaining buildings according to environmentally sustainable principles, such as LEED certification. Example: The Reserve Bank of New Zealand incorporates green building standards in its new office developments to minimize environmental impact.

Apart from these many others praiseworthy initiatives such as waste management, carbon footprint measurement, water recycling etc. have already been taken by several monetary authorities

### 3.4.2 Social Practices

**Diversity and Inclusion:** Central banks prioritize diversity in hiring and foster inclusive workplace cultures.

Example: The Bank of England has set targets to achieve gender parity in leadership roles by February 2028, supported by diverse initiatives.[22]

**Employee Well-being:** Supporting staff through wellness programs, professional development opportunities, and ensuring fair labor practices.

**Community Engagement:** Engaging with local communities through outreach programs and initiatives that promote financial literacy and inclusion.

Example: The Bank of Canada collaborates with community organizations to support financial education programs for underserved populations [23]

**Supplier Diversity:** Promoting diversity among suppliers and contractors to support economic inclusion.

### 3.4.3 Governance Practices:

**Transparency and Accountability:** Maintaining high standards of ethical conduct, transparency in operations, and accountability to stakeholders.

**Risk Management:** Implementing robust risk management frameworks to identify, assess, and mitigate risks related to ESG factors.

**Ethical Investment Practices:** Integrating ESG considerations into investment decisions, aligning investments with sustainability goals.

**Stakeholder Engagement:** Actively engaging with stakeholders, including shareholders, employees, regulators, and the public, to address ESG issues transparently.

**Training and Awareness:** Providing training and awareness programs for employees on ESG issues and their importance in central banking operations. Several Central Banks introduced in house training and capacity building programs to strengthen their human resources capabilities to facilitate policy to combat climate change risk in financial markets. For example, the Bank of New Zealand committed to provide training on climate risk to all of its staff by the end of 2021[24].

## 5. Central Bank Case Study on Sustainability Initiatives

This section presents detailed case studies of central banks that have successfully integrated sustainability into their operations and policies. Examples include:

### 5.1 Bangladesh Bank: Greening the Banking Sector

Bangladesh Bank (BB) has been proactive in incorporating sustainability into its financial system, spearheading initiatives to green the banking sector. Here are some of the key efforts and strategies undertaken by BB:

#### 5.1.1 Policy Formulation and Directives

Green Banking Policy Guidelines: In 2011, Bangladesh Bank issued guidelines for green banking to encourage banks to integrate environmental risk management into their credit policies.

Environmental Risk Management (ERM) Guidelines: These guidelines, issued in 2011, aim to help banks assess environmental risks as part of their credit risk assessment process.

#### 5.1.2 Promoting Green Financing

**Refinance Schemes:** BB has introduced several refinance schemes at preferential rates to promote green products and projects. These include financing for Green Products, Green Transformation Fund (GTF), Brick Kiln Efficiency Improvement Project etc [25].

**Incentives for Green Projects:** BB provides incentives such as lower interest rates and extended loan tenures for projects with a positive environmental impact.

#### 5.1.3 Sustainable Finance Department

Bangladesh established a dedicated Sustainable Finance Department to oversee and implement green banking and sustainable finance initiatives. This department monitors compliance with green banking guidelines and supports capacity building within the banking sector.

#### 5.1.4 Green Reporting and Disclosure

**Quarterly Reporting:** Schedule banks and financial institutions are required to submit annual reports on their green banking activities, including the number and volume of green projects financed [26].

**Transparency and Accountability:** These reports are aimed at ensuring transparency and accountability in the sector's efforts towards sustainability.

#### 5.1.5 Awareness and Capacity Building

**Training and Workshops**:

Public Awareness Campaigns: Efforts are made to educate the public on the benefits of green financing and sustainable banking practices.

## 5.1 6 Collaboration and Partnerships

**National and International Collaboration:** BB collaborates with national and international organizations, including the Sustainable and Renewable Energy Development Authority (SREDA), to promote sustainable finance.

**Global Initiatives:** Bangladesh Bank also participate in global initiatives such as the Sustainable Banking Network (SBN) to align local practices with international standards.

## 5.1.7 Monitoring and Evaluation

**Regular Monitoring:** BB regularly monitors the progress of banks in implementing green banking policies and achieving their sustainability targets.

**Impact Assessment:** Evaluations are conducted to assess the impact of green financing on the environment and the economy.

Bangladesh Bank's comprehensive approach to greening the banking sector has set a benchmark in sustainable finance. By integrating environmental considerations into the financial system, BB aims to promote sustainable economic growth while mitigating environmental risks. Through continued efforts in policy formulation, capacity building, and innovative financing solutions, BB is driving the transition towards a more sustainable and resilient banking sector in Bangladesh.

## 5.2 The People's Bank of China: Journey Towards NetZero

The People's Bank of China (PBOC) has been at the forefront of integrating sustainability into its financial policies, aiming to support China's ambitious goal of achieving net-zero carbon emissions by 2060. Here are some of the key steps and initiatives undertaken by PBOC in this journey:

### 5.2.1 Policy Framework and Guidelines

- **Green Finance Guidelines**: PBOC has issued comprehensive guidelines to promote green finance, encouraging banks and financial institutions to develop and invest in green projects.
- **Green Bond Standards**: Establishment of green bond standards to ensure transparency and credibility in the issuance of green bonds.

### 5.2.2 Green Financing Initiatives

- **Green Credit Policy**: Banks are incentivized to provide credit for projects that are environmentally sustainable. This includes preferential interest rates for green projects and stricter credit assessments for high-polluting industries.
- **Green Bond Market**: China has become one of the largest issuers of green bonds globally, with PBOC playing a crucial role in developing a robust green bond market.

### 5.2.3 Carbon Market Development

- **National Carbon Trading System**: PBOC supports the development of China's national carbon trading market, which aims to cap and reduce carbon emissions from key industries.
- **Carbon Pricing Mechanisms**: Implementation of carbon pricing mechanisms to encourage industries to adopt low-carbon technologies and practices.

### 5.2.4 Monetary Policy Adjustments

- **Green Monetary Policy Tools**: Introduction of green monetary policy tools, such as green re-lending facilities and green reserve requirements, to support green finance.
- **Interest Rate Incentives**: Offering lower interest rates for green loans and projects to make them more attractive for investors and developers.

### 5.2.5 International Collaboration

- **Belt and Road Initiative (BRI)**: Promoting green finance within the Belt and Road Initiative, ensuring that infrastructure projects under the BRI adhere to environmental sustainability standards.
- **Partnerships with International Organizations**: Collaborating with international organizations like the United Nations, International Monetary Fund (IMF), and World Bank to align with global sustainability standards and practices.

### 5.2.6 Research and Development

- **Green Finance Research**: Establishment of dedicated research institutions to study and promote green finance practices and policies.
- **Climate Risk Assessment**: Conducting research on climate-related financial risks and developing tools to assess and mitigate these risks within the financial system.

### 5.2.7 Regulatory and Supervisory Measures

- **Green Finance Regulations**: Implementation of regulations to ensure that financial institutions incorporate environmental risk management into their operations.
- **Supervisory Framework**: Developing a supervisory framework to monitor and evaluate the performance of financial institutions in terms of their green finance activities.

These efforts not only contribute to China's environmental goals but also set an example for other countries to follow in integrating sustainability into their financial systems.

### 5.3 The European Central Bank (ECB): Introduction of Climate Stress Testing

The European Central Bank (ECB) has recognized the significant impact that climate change can have on financial stability and has taken steps to incorporate climate risk into its supervisory and financial stability frameworks. One of the key measures introduced by the ECB is climate stress testing.

What is Climate Stress Testing?

Climate stress testing is a tool used to assess the resilience of financial institutions to climate-related risks. It evaluates how banks and other financial entities would perform under various climate scenarios, including those involving severe climate impacts. The aim is to understand potential vulnerabilities and ensure that financial institutions can withstand climate-related shocks. The ECB climate risk stress testing framework incorporates a detailed methodology for assessing the effects of climate scenarios on the financial institutions over a 30 years period in future [27].

Objectives of ECB's Climate Stress Testing

**Assess Financial Stability:** To gauge the impact of climate-related risks on the overall stability of the financial system.

**Inform Supervision:** To provide insights that can be used to enhance the supervision of banks and ensure they manage climate risks appropriately.

**Promote Transparency:** To increase transparency regarding the climate-related risks faced by financial institutions.

**Guide Policy:** To support the development of policies aimed at mitigating climate risks within the financial sector.

Innovative Components of ECB's Climate Stress Testing

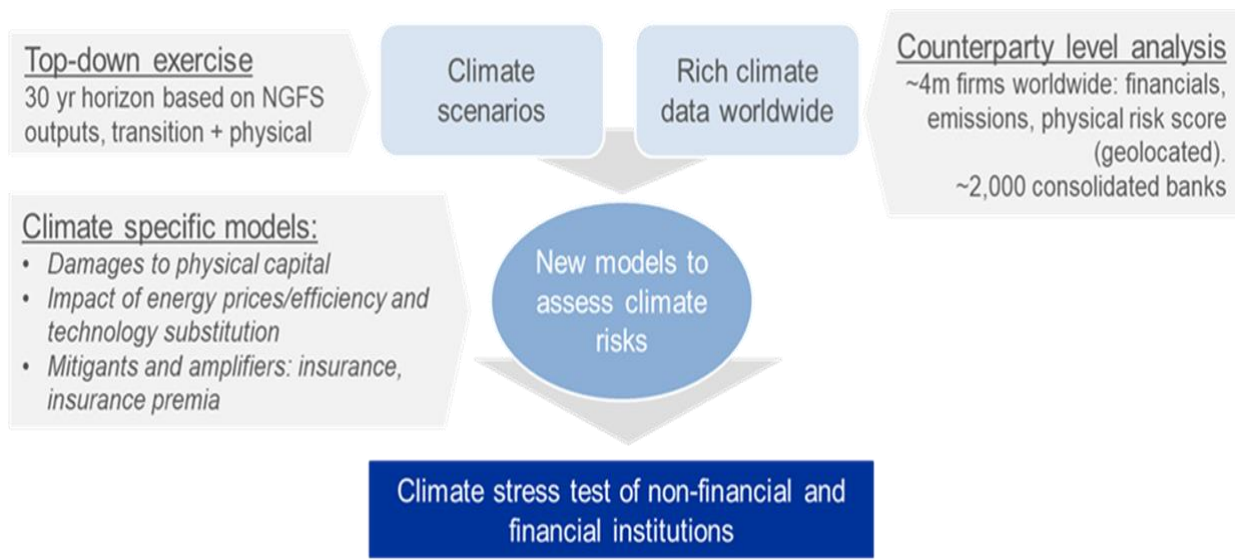

Figure 2: Element of the ECB Climate Stress Test. Source: ECB

**Centralized, Top Down Exercise:** The ECB developed the framework based on data, assumptions and models which applied to all financials institutions that are part of the sample. For that reason, the model is different from the bottom up exercise where banks their exposure to climate related risk and adaptability to the risks [27].

**Climate Specific Scenario Analysis:** This framework took accounts for both transition and physical risk of financial institutions for climate change whereas most of the model heavily focused on transition risk.

**Counterparty Level Analysis:** The framework is based on robust dataset that consolidated ECB analytical Credit Datasets, ECB Securities Holding Statistics and other relevant data sources to accurately map bank's NFC counterparties with carbon emission record, physical risk exposures. [27]

**Climate Specific Model:** Firms' risk levels depended on GHG emissions, energy mix, innovation, and geographical location. The methodology enabled dynamic modeling of credit and market risks, with potential extensions to other sectors and regulatory assessments.

## Implementation Timeline

The ECB has been progressively implementing climate stress testing. Key milestones include:

2020: The ECB published its guide on climate-related and environmental risks, outlining expectations for banks.

2021: The ECB launched its first economy-wide climate stress test.

2022: The ECB conducted its first supervisory climate stress test focused on individual banks.

Ongoing: The ECB continues to refine its approach, with regular updates and further integration of climate risk into its supervisory processes.

## Challenges and Considerations

Data Gaps: Accurate and comprehensive data is crucial for effective stress testing. The ECB works to address data gaps and improve data quality.

Modeling Complexity: Climate stress testing involves complex modeling and scenario analysis, requiring advanced tools and methodologies.

Interdisciplinary Approach: Combining expertise from climate science, economics, and finance is essential to develop robust stress tests.

The ECB's introduction of climate stress testing is a critical step in ensuring the resilience of the European financial system to climate-related risks. By proactively identifying and addressing these risks, the ECB aims to safeguard financial stability and promote sustainable finance.

## 5.4 Federal Reserve System: Inauguration of Climate Scenario Analys

The Federal Reserve System (Fed), recognizing the growing impact of climate change on the economy and financial system, has taken steps to incorporate climate-related risks into its supervisory framework. One significant initiative is the introduction of climate scenario analysis.

### What is Climate Scenario Analysis?

Climate scenario analysis involves examining how different potential future climate pathways could impact the financial system. It allows policymakers and financial institutions to evaluate risks and develop strategies to mitigate them. Unlike traditional stress testing, which focuses on short-term shocks, climate scenario analysis often looks at long-term impacts.

### Objectives of the Fed's Climate Scenario Analysis

**Assess Financial Stability:** To evaluate the potential impact of climate-related risks on the stability of the financial system.

**Inform Supervision and Regulation:** To integrate climate risk into the supervisory process, ensuring that financial institutions are prepared to manage these risks.

**Enhance Understanding:** To improve the understanding of climate-related financial risks among policymakers and financial institutions.

**Support Policy Development**: To provide insights that inform the development of policies aimed at addressing climate risks.

### Key Components of the Fed's Climate Scenario Analysis

**Scenario Development:** Crafting a range of plausible climate scenarios, considering factors such as policy responses, technological changes, and physical impacts of climate change.

Data Collection and Analysis: Gathering data on financial institutions' exposures to climate-sensitive sectors and geographies and analyzing how these exposures might be affected under different scenarios.

Risk Assessment: Evaluating both physical risks (e.g., impacts from extreme weather events) and transition risks (e.g., risks associated with the transition to a low-carbon economy).

Reporting and Feedback: Communicating findings to financial institutions and providing guidance on managing identified risks.

### Implementation Timeline

The Fed's climate scenario analysis is part of a broader effort to address climate-related financial risks. Key steps include:

**2020:** The Fed joined the Network for Greening the Financial System (NGFS), a coalition of central banks and supervisors focused on addressing climate risks [28].

**2021:** The Fed established a Supervision Climate Committee to enhance its understanding and supervisory approach to climate-related risks [29].

**2022:** The Fed announced plans to conduct pilot climate scenario analyses with a select group of large financial institutions.

**2023:** The Federal Reserve conducted a pilot climate scenario analysis (CSA) exercise to understand the climate risk management practices and challenges faced by large banking organizations. The exercise aimed to improve the capacity of these organizations and supervisors to identify, estimate, monitor, and manage climate-related financial risks [30].

**Ongoing:** The Fed continues to refine its methodologies and expand its climate scenario analysis efforts. Recently the Federal Reserve Bank of New York along with Enterprise Community Partners announced publication of "What's Possible: Investing NOW for Prosperous, Sustainable Neighborhoods." This book consolidated essays from leading climate finance and community experts [31].

### Challenges and Considerations

**Data Limitations:** Obtaining reliable and comprehensive data on climate risks remains a challenge. The Fed is working to improve data quality and availability.

**Modeling Complexity:** Developing robust models that accurately capture the long-term and uncertain nature of climate risks requires advanced techniques and interdisciplinary collaboration.

**Regulatory Integration:** Integrating climate scenario analysis into existing regulatory frameworks involves balancing new insights with established practices.

### Future Directions

**Enhanced Collaboration:** The Fed is likely to increase collaboration with other central banks, financial institutions, and stakeholders to share best practices and improve methodologies.

**Broader Scope:** Expanding climate scenario analysis to include a wider range of financial institutions and more diverse scenarios.

**Ongoing Refinement:** Continuously updating and improving scenario analysis tools and techniques to reflect the latest scientific and economic research.

The Federal Reserve System's inauguration of climate scenario analysis represents a significant step in integrating climate-related risks into its supervisory framework. By proactively assessing these risks, the Fed aims to enhance the resilience of the financial system and support the transition to a more sustainable economy.

### 5.5 The Central Bank of Kenya: Higher Capital Requirements for Climate Risk

The Central Bank of Kenya (CBK) has been proactive in addressing climate-related financial risks, recognizing the potential impact of climate change on the stability of the financial system. One of the key measures implemented by the CBK is the introduction of higher capital requirements for climate risk. This approach aims to ensure that banks maintain sufficient capital buffers to withstand potential losses arising from climate-related risks.

Objectives of Higher Capital Requirements for Climate Risk

Enhance Resilience: To increase the resilience of banks to climate-related financial shocks by requiring them to hold additional capital.

Promote Risk Management: To encourage banks to integrate climate risk into their risk management frameworks and decision-making processes.

Safeguard Financial Stability: To protect the stability of the financial system from the potential destabilizing effects of climate change.

Support Sustainable Finance: To incentivize banks to finance projects and initiatives that support sustainability and climate resilience.

Key Components of the CBK's Approach

Risk Assessment Framework: Developing a comprehensive framework for assessing climate-related risks, including both physical risks (e.g., natural disasters) and transition risks (e.g., regulatory changes).

Capital Requirement Adjustment: Setting higher capital requirements for banks with significant exposures to climate-sensitive sectors or activities, ensuring they have adequate buffers to absorb potential losses.

Supervisory Review: Incorporating climate risk assessments into the regular supervisory review and evaluation process, with a focus on banks' risk management practices and capital adequacy.

Reporting and Disclosure: Mandating enhanced reporting and disclosure of climate-related risks and capital adequacy, promoting transparency and accountability.

Implementation Timeline

The CBK has taken a phased approach to implementing higher capital requirements for climate risk, with key milestones including:

2020: The CBK issued guidelines on climate-related risk management, outlining expectations for banks to assess and manage climate risks.

2021: The CBK conducted a survey of banks to understand their exposure to climate risks and their preparedness in managing these risks.

2022: The CBK introduced higher capital requirements for banks with significant climate risk exposures, providing a transition period for banks to comply.

Ongoing: CBK continues to monitor banks' compliance with the new requirements and refine its approach based on emerging best practices and feedback from the industry.

Challenges and Considerations

Data Availability: Ensuring access to reliable and comprehensive data on climate risks is crucial for effective implementation. CBK is working to improve data collection and sharing.

Risk Differentiation: Differentiating between banks based on their specific climate risk exposures and adjusting capital requirements accordingly.

Capacity Building: Enhancing the capacity of both the CBK and banks to understand and manage climate-related risks through training and technical assistance.

Market Impact: Considering the potential impact of higher capital requirements on the availability and cost of credit, particularly for climate-sensitive sectors.

Future Directions

Continuous Improvement: The CBK will continue to refine its climate risk assessment methodologies and capital requirement frameworks based on new research and industry developments.

International Collaboration: Engaging with international regulatory bodies and other central banks to share best practices and harmonize approaches to climate risk regulation.

Stakeholder Engagement: Collaborating with banks, industry associations, and other stakeholders to ensure effective implementation and address any challenges that arise.

Conclusion

The Central Bank of Kenya's introduction of higher capital requirements for climate risk is a proactive step towards enhancing the resilience of the financial system to climate-related shocks. By requiring banks to hold additional capital, the CBK aims to promote sound risk management practices and support the transition to a sustainable and climate-resilient economy.

## 5.6 The Reserve Bank of New Zealand: New Guidance for climate risk

The Reserve Bank of New Zealand (RBNZ) has recognized the critical importance of addressing climate-related financial risks and has introduced new guidance to help financial institutions manage these risks more effectively. This initiative is part of RBNZ's broader commitment to promoting financial stability and sustainability.

Objectives of the RBNZ's New Guidance for Climate Risk

Enhance Risk Management: To improve the capacity of financial institutions to identify, assess, and manage climate-related risks.

Promote Transparency: To increase transparency and disclosure of climate-related financial risks.

Support Regulatory Oversight: To provide a clear framework for supervisory oversight of climate risks.

Encourage Sustainable Practices: To incentivize financial institutions to support sustainable and climate-resilient investments.

Key Components of the RBNZ's Guidance

Risk Identification and Assessment: Financial institutions are encouraged to systematically identify and assess climate-related risks, including both physical risks (e.g., extreme weather events) and transition risks (e.g., policy changes related to climate mitigation).

Integration into Risk Management: Institutions are advised to integrate climate-related risks into their existing risk management frameworks, ensuring these risks are considered alongside traditional financial risks.

Scenario Analysis and Stress Testing: The guidance recommends that financial institutions use scenario analysis and stress testing to evaluate their exposure to climate risks under different future scenarios.

Governance and Oversight: Institutions should establish robust governance structures to oversee the management of climate-related risks, including board-level responsibility and appropriate risk committees.

Disclosure and Reporting: The guidance emphasizes the importance of transparent reporting and disclosure of climate-related risks, aligned with international frameworks such as the Task Force on Climate-related Financial Disclosures (TCFD).

Implementation Timeline

The RBNZ has laid out a phased approach to implementing its new guidance on climate risk:

2020: The RBNZ published a discussion paper outlining its approach to incorporating climate-related risks into its regulatory framework.

2021: The RBNZ engaged with stakeholders, including banks and insurers, to gather feedback and refine its guidance.

2022: The RBNZ issued its final guidance on managing climate-related risks, providing financial institutions with detailed expectations and best practices.

Ongoing: The RBNZ continues to monitor the implementation of the guidance and engage with financial institutions to ensure compliance and address any emerging challenges.

Challenges and Considerations

Data Gaps: Access to high-quality, granular data on climate risks remains a challenge. The RBNZ is working to improve data availability and encourage data sharing among institutions.

Methodological Complexity: Developing robust methodologies for assessing and managing climate risks, including scenario analysis and stress testing, requires ongoing refinement and capacity building.

Regulatory Coordination: Ensuring alignment with international standards and practices while tailoring the guidance to the specific context of New Zealand's financial system.

Stakeholder Engagement: Continuously engaging with financial institutions and other stakeholders to address concerns and incorporate feedback into the regulatory framework.

Future Directions

Enhanced Monitoring: The RBNZ plans to enhance its monitoring of climate-related risks and the effectiveness of the guidance through regular supervisory reviews and thematic assessments.

Capacity Building: Providing training and resources to financial institutions to strengthen their capabilities in managing climate risks.

International Collaboration: Collaborating with other central banks, regulators, and international organizations to share best practices and stay abreast of global developments in climate risk management.

Policy Integration: Exploring ways to further integrate climate considerations into monetary policy, financial stability assessments, and other areas of the RBNZ's mandate.

Conclusion

The Reserve Bank of New Zealand's new guidance for climate risk represents a significant step towards integrating climate considerations into the financial system. By providing a clear framework for managing climate-related risks, the RBNZ aims to enhance the resilience of financial institutions and support the transition to a sustainable and climate-resilient economy.

## 6. Criticism on Central Bank integration of Sustainability

Central banks are at the forefront of the transition towards a sustainable economy. By leveraging their unique position and influence, they can drive significant progress in sustainable finance. Their efforts are crucial in achieving global sustainability targets and addressing the pressing environmental and social challenges of our time. The imperative for central banks to support sustainable development is clear, and their role in shaping a sustainable financial system is indispensable.

While central bank initiatives in sustainability, including the integration of climate-related risks into their regulatory frameworks and financial stability assessments, have generally been welcomed, they have also faced several criticisms. Here are some of the main points of contention:

1. Mandate Overreach

Criticism:

Scope of Mandate: Critics argue that central banks are overstepping their traditional mandates, which primarily focus on monetary policy and financial stability. They suggest that addressing climate change and sustainability issues should fall under the purview of elected governments and specialized environmental agencies, not central banks.

Counterargument:

Financial Stability: Proponents assert that climate risks pose a significant threat to financial stability, which is within the core mandate of central banks. Ignoring these risks could lead to severe economic disruptions.

2. Effectiveness and Expertise

Criticism:

Effectiveness: Some argue that central banks lack the necessary expertise and tools to effectively address complex and multifaceted climate-related risks. They worry that central banks' actions might not significantly impact climate change outcomes.

Focus on Core Competencies: Critics suggest that central banks should focus on their core competencies, such as controlling inflation and ensuring financial stability, rather than taking on additional roles.

Counterargument:

Capacity Building: Central banks are investing in capacity building and collaborating with experts in climate science and sustainability to enhance their understanding and effectiveness in this area.

Global Cooperation: Through initiatives like the Network for Greening the Financial System (NGFS), central banks are working together to develop best practices and share knowledge, improving their collective ability to address climate risks.

3. Unintended Consequences

Criticism:

Market Distortions: Imposing higher capital requirements for climate risk or mandating specific disclosures could lead to market distortions, potentially driving up the cost of credit and affecting the availability of financing for certain sectors.

Regulatory Burden: There is concern that new regulations could impose additional compliance costs on financial institutions, particularly smaller banks, which might struggle to meet these requirements.

Counterargument:

Risk Mitigation: The measures are designed to mitigate long-term risks that could have far more severe economic consequences if left unaddressed.

Phased Implementation: Central banks often use a phased approach to implementation, providing financial institutions with time to adapt and ensuring that regulatory measures are proportionate and targeted.

4. Political Neutrality

Criticism:

Political Pressure: Involvement in sustainability issues could expose central banks to political pressure and compromise their independence. Critics worry that central banks might be co-opted into advancing political agendas, undermining their credibility.

Counterargument:

Objective Risk Management: Central banks emphasize that their actions are based on objective risk management considerations and are aimed at ensuring financial stability. They argue that addressing climate risks is a matter of prudence, not politics.

5. Global Inequities

Criticism:

Developing Economies: Central banks in developing countries might face greater challenges in implementing sustainability initiatives due to limited resources and differing economic priorities. There is concern that global standards might not adequately consider these disparities.

Counterargument:

Tailored Approaches: Central banks can tailor their approaches to reflect the specific contexts and needs of their economies, and international frameworks often provide flexibility to accommodate such differences.

Support and Collaboration: International organizations and developed economies can provide support and resources to help developing countries build their capacities in managing climate-related financial risks.

# 7. Best Practices, Challenges, Future Directions

## 7.1 Best Practices for Central Banks in scale up sustainable finance

Based on the case studies, this section summarizes best practices for central banks to advance sustainable finance. Key recommendations include:

Integrating climate-related risks into monetary policy and economic forecasting. This involves developing new models and tools to understand the macroeconomic implications of climate change.

Promoting transparency and disclosure of climate-related risks by financial institutions. This can be achieved through regulatory requirements and guidelines that ensure consistent and comparable disclosures.

Supporting the development of green financial products and markets. Central banks can play a catalytic role by providing liquidity to green markets and setting standards for green financial instruments.

Collaborating with governments and international institutions to align financial systems with sustainability goals. Effective collaboration ensures that policies are coherent and mutually reinforced, enhancing their overall impact.

## 7.2 Challenges and Future Directions

### 7.2.1 Challenges in Advancing Sustainable Finance

Despite the progress made, central banks face several challenges in advancing sustainable finance. These challenges include:(Governors et al., n.d.)

Limited data and methodologies for assessing climate-related risks. Accurate and comprehensive data is crucial for effective risk assessment and policymaking.

Potential conflicts between traditional mandates (e.g., price stability) and sustainability goals. Central banks need to balance their traditional objectives with the emerging imperative to support sustainable development.

The need for greater coordination and cooperation among central banks, governments, and financial institutions. Effective collaboration is essential for addressing global sustainability challenges.

### 7.2.2 Future Directions for Central Banks

Looking ahead, central banks can play an even more significant role in promoting sustainable finance. Future directions include:

Enhancing climate risk assessment tools and methodologies. This involves developing more sophisticated models and improving data availability.

Expanding the scope of green financial instruments and markets. Central banks can support the development of new green financial products and enhance the liquidity of green markets.

Strengthening international cooperation to address global sustainability challenges. Central banks can work together to develop global standards and share best practices.

Continuously adapting and evolving their policies and practices to support sustainable development. This requires ongoing innovation and a willingness to learn from experience.

## 8 Implications for Policy and Practice

The findings of this chapter have significant implications for policymakers and financial institutions. Central banks need to balance their traditional mandates with the emerging necessity to support sustainable development. This requires a holistic approach that integrates sustainability into all aspects of central banking operations and policies.

While central bank initiatives in sustainability are essential for addressing the financial risks posed by climate change, they are not without controversy. Balancing the benefits of proactive risk management with the potential drawbacks requires careful consideration, ongoing dialogue with stakeholders, and a commitment to evolving approaches based on emerging evidence and best practices.